\documentclass[11pt]{article}

\usepackage{amssymb}
\usepackage{citesort}

\title{Properties}

\begin{document}

\begin{titlepage}

\begin{flushright}
                LPT Orsay 03/09\\
                February 2003  \\
                hep-th/0302211
\end{flushright}
\bigskip

\begin{center}

{\LARGE \bf Non-critical string \\
Liouville theory \\
and \\
geometric bootstrap hypothesis \\}

\end{center}
\bigskip

\begin{center}
    {\bf
    Leszek Hadasz}\footnote{e-mail: hadasz@th.u-psud.fr} \\
    Laboratoire de Physique Th\`{e}orique, B{\^a}t. 210,\\
    Universit{\'e} Paris-Sud,  91405 Orsay, France \\
        {\em and }\\

    Jagiellonian University, 
    Reymonta 4, 30-059 Krak\'ow, Poland \\
\vskip 5mm
    {\bf
    Zbigniew Jask\'{o}lski}\footnote{e-mail: Z.Jaskolski@if.uz.zgora.pl }\\
    Physics Institute\\
    University of Zielona G\'{o}ra\\
    ul. Szafrana 4a,
    65-069 Zielona G\'{o}ra, Poland

\end{center}

\vskip 1cm

\begin{abstract}
The applications of the existing Liouville theories for the
description of the longitudinal dynamics of 
non-critical Nambu-Goto string are analyzed.
We show that the recently developed DOZZ solution 
to the Liouville theory leads to the 
cut singularities in tree string amplitudes.
We propose a new version of the Polyakov geometric approach to
Liouville theory and formulate its basic consistency condition --- the
geometric bootstrap equation. 
Also in this approach the tree amplitudes develop cut singularieties.
\end{abstract}
\bigskip\bigskip

\thispagestyle{empty}
\end{titlepage}

\section{Introduction}
\setcounter{equation}{0}
It has been well known since early days of string theory that the
covariant quantization of the free Nambu-Goto string
\cite{brower72}  leads in non-critical dimensions ($1<d<25$) to
free quantum models with longitudinal excitations. However,
in spite of numerous attempts no consistent theory of the longitudinal
dynamics has been found. 
Recently the problem was reconsidered in \cite{Hadasz:2002gk}.
It was shown that if a consistent theory of the longitudinal dynamics exists
it does not satisfy  all the axioms of the standard CFT.  
This explains the failure of previous approaches
and makes it difficult to construct a viable alternative.

In the present paper we analyze whether the existing models 
of quantum Liouville theory provide an adequate description of the longitudinal
dynamics. 
We restrict ourselves to non-critical strings 
with the space of
asymptotic free states coinciding with the space of states of the free
non-critical Nambu-Goto string ($1<d<25$ )\cite{dahaja98}.
Although the model is inconsistent due to the presence of a
tachyon it is still a good starting point for the analysis of the longitudinal dynamics. 
The experience with the critical string shows that the tachyon problem  
is to large extend independent of the consistency of the string perturbation 
expansion, the covariance and the unitarity issues.
Moreover, the
standard GSO mechanism \cite{Gliozzi:1976qd} can be used  to remove the tachyon 
in the free non-critical RNS string ($1<d<9$) \cite{Hasiewicz:1999nr}. We may hope therefore that 
the basic properties of the longitudinal dynamics derived in the context of the 
noncritical Nambu-Goto string should hold  in 
more realistic models as well.

The relevance of the quantum Liouville theory 
 for the proper description  of the longitudinal string excitations has been known 
since the celebrated papers by A.~Polyakov on conformal anomaly \cite{poly81,poly81b}. 
It was conjectured some time ago by  A. and Al. Zamolodchikov \cite{Zamolodchikov:1995aa}
that the continuum spectrum
along with the 3-point functions proposed by Otto and Dorn \cite{Dorn:1994xn}
 satisfy the bootstrap
consistency conditions of standard (in the sense of BPZ \cite{Belavin:1984vu}) CFT. 
In the weak coupling regime $c>25$ this conjecture was recently proved by Ponsot and Teschner
\cite{Ponsot:1999uf,Ponsot:2000mt}.
 It is believed to hold in the strong coupling
regime $1<c<25$ by the analytic continuation argument \cite{Teschner:2001rv}.
Among many recent applications of 
this elegant solution  let us only mention the
1+1 dimensional string models 
\cite{McGreevy:2003kb,Klebanov:2003km,McGreevy:2003ep,Alexandrov:2003nn} 
(where its
predictions are confirmed by the matrix model results 
\cite{gm,DiFrancesco:1993nw}),
quantization of the Teichm\"uller space of Riemann surfaces 
\cite{Verlinde:1989ua,Teschner:2002vx,Teschner:2003em},
relations with WZNW models, 
\cite{Gawedzki:1991yu,Teschner:1999ug} 
and string theory on the $AdS_3$ space
\cite{Maldacena:2000hw,Giveon:2001up}.

The continuous spectrum plays an essential role
in all the applications mentioned above. On the other hand 
it is the main obstacle in applying the theory to the non-critical
Nambu-Goto string  where the longituinal sector consists 
of a single conformal family.
The straightforward application of the DOZZ theory
yields 
consistent perturbation expansion and unitarity. 
The price one has to pay is the extension of the spectrum of external
string states far beyond the spectrum of the free non-critical Nambu-Goto.
This  leads to the continuous family of intercepts and is not acceptable on physical grounds.
A possible way out is to restrict ourselves to  the external states from the single
conformal family of the longitudinal sector. In this case however the tree amplitudes exhibit cut singularities.

There exist another approach to the quantum Liouville theory originally
proposed by 
A.~Polyakov  \cite{Polyakov82} and developed by Takhtajan 
\cite{Takhtajan:yk,Takhtajan:1993vt,Takhtajan:zi,Takhtajan:1994vt,Takhtajan:1995fd}.
In this geometric approach
the correlation functions
are defined  in terms of path integral over conformal class of Riemannian metrics
with prescribed  singularities at the punctures.
The semi-classical results obtained in the case of  parabolic singularities \cite{Takhtajan:yk,Takhtajan:1993vt,Takhtajan:zi,Takhtajan:1994vt}
are in perfect agreement with the general properties of the
longitudinal dynamics derived in \cite{Hadasz:2002gk}. 
An additional support for the  geometrical approach
comes from  the fact that some of its geometric
predictions  can be rigorously proved
(\cite{Takhtajan:zi,Takhtajan:1994vt}, and references therein).
In spite of considerable achievements 
the geometric approach  is not yet capable to
produce the puncture correlators. For this reason both the
relation to the DOZZ theory \cite{Takhtajan:1995fd} and the application to
longitudinal dynamics remain open problems. 

As a step toward calculation of puncture correlators we propose to suplement 
the geometric approach by  a  new dynamical principle - the geometric bootstrap
equation. We are not able to prove that structure constants satisfying 
this equation exist. However assuming their analytic dependence on the conformal weights
one can show that the 4-point correlator leads to the cut singularities of string amplitudes
which coincides with the result obtained from the DOZZ solution.

With the standard structure of the string perturbation expansion
\cite{Veneziano:dr} assumed, 
 the cut singularities of tree amplitudes 
indicate the continuous spectrum of intercepts. 
This is a strong evidence that either the consistent
non-critical Nambu-Goto string does not exist at all 
or the existing Liouville models does not  provide
an adequate description of the longitudinal dynamics.
The third possibility that the 
perturbation expansion of non-critical string 
 does not have the structure
known from the critical string theory
is a difficult open
problem going beyond the scope of the present paper.

The paper is organized as follows. In Sect.2  we summarize those properties of the longitudinal
dynamics which are necessary for the Lorentz covariance of the tree light-cone amplitudes
of non-critical Nambu-Goto string. 
In Sect.3 we demonstrate that  the  solution provided by the DOZZ theory 
leads to the tree non-critical strings 
with cut singularities. 
In Sect.4, using general properties of the geometric approach to the Liouville theory we 
formulate the geometric bootstrap equation. Using the results of Wolf and Wolpert on
the asymptotic behaviour of the geodesic length \cite{Wolf} we show that the singularities 
of string amplitude are of the same type as in the case of the DOZZ solution.
Finally Sect.5 contains conclusions and brief discussion of open problems.

\section{General properties of the longitudinal dynamics}
\setcounter{equation}{0}

In the light-cone formulation string amplitudes 
are constructed in terms of 2-dim field theory on  light-cone diagrams describing time ordered sequences of 
elementary splitting and 
joining processes \cite{Mandelstam73,Mandelstam74,Mandelstam86}. 
The problem of introducing interactions can be seen as a problem of extending 
the theory from the cylinder, where it is completely determined by the
free string theory at hand, 
to an arbitrary light-cone diagram. 
In the case of the Nambu--Goto non-critical string the extension in the transverse sector
is  given by the tensor product of $d-2$ copies of the scalar CFT. 

According to our assumptions concerning the free theory
the space of states in the longitudinal sector 
is a tensor product of single left and single right Verma
module ${\cal V}_h\otimes \bar {\cal V}_{\bar h}$. 
The central charges $c=\bar c$
and the highest weights $h=\bar h$ are related to the dimension of the target space by
\cite{dahaja98} :
$$
c= 1+6Q^2 \;=\; 26-d \ ,\;\;\;\;h\;=\;{Q^2\over 4}\;=\; 
{25-d\over 24}\ .
$$
The only ground state 
$|\,0\,\rangle=\omega\otimes\bar{\omega}$  
is not $PSL(2,\mathbb{C})$-invariant and 
the energy momentum-tensor\footnote{The antyholomorphic counterparts of
the formulae are assumed.} 
\begin{eqnarray}
\label{tensor}
{T(z)}
&=&
\sum\limits_{n}L_nz^{-n-2}
\end{eqnarray}
is singular in the limit $z\to 0,$
$$
T(z)
|\,0\,\rangle
= {Q^2 \over 4 z^2}
|\,0\,\rangle
+ {1\over z}L_{-1}
|\,0\,\rangle
+ {\rm regular \; terms} \ .
$$
The ground state
applied to the free end of a semi-infinite cylinder thus
corresponds to a puncture on the complex plane with prescribed singularity structure of
the energy--momentum tensor at it.

In the case of tree light-cone diagrams with $N$ arbitrary external states
the longitudinal sector is described by correlation
functions of appropriate number of the energy--momentum tensor
insertions on the Riemann sphere $S^2(z_1,\dots,z_N)$ with $N$
punctures:
\begin{equation}
\label{badnotation}
\left\langle\, \prod_j T(w_j)
\prod_k \widetilde T(\bar w_k)
\,\right\rangle_{S^2(z_1,\dots,z_N)} \ .
\end{equation}
Such formulation may seem strange from the point of view of the standard CFT.
Indeed if we had assumed the operator--state correspondence we could have replaced
the correlation functions on $N$-punctured sphere  by 
correlation functions on the sphere with no punctures but with additional local operator
insertions. As we shall see
the space of states in the longitudinal sector
is ``too small'' to accommodate such construction. 
Even though the operator--state correspondence is not assumed it is convenient to replace  
 the adequate but clumsy notation
(\ref{badnotation}) by 
\begin{equation}
\label{Lcorrelators}
\left\langle \,\prod_j T(w_j)
\prod_k \widetilde T(\bar w_k)
\prod\limits_{r=1}^N P(z_r,\bar z_r)\,\right\rangle \ .
\end{equation}
The properties of the longitudinal sector which  are  necessary 
for the Lorentz covariance of tree string amplitudes can be summarized as follows \cite{Hadasz:2002gk}\medskip

\noindent {\bf I}
{\it 
The conformal anomaly 
has its universal form given by the (regularized) Liouville action 
on the punctured sphere and
depends on the central charge $c=1+6Q^2$ in the standard way.}
\medskip

\noindent {\bf II}
{\it 
For all correlation functions on $N$-punctured spheres 
the operator--operator product expansion (OOPE) and
the operator--puncture product expansion (OPPE) hold:}
\begin{eqnarray}
\label{TTOPE}
T(w)T(z)
&=&{1+6Q^2\over 2(w-z)^4 }
+ {2 \over (w-z)^2}T(z)
+{ 1\over w-z}\partial T(z) + \dots \ , \\
\label{TPOPE}
T(w) P(z, \bar z) &=&
{Q^2\over 4(w-z)^2}P(z, \bar z)
+{ 1 \over w-z}\partial P(z, \bar z) + \dots \ .
\end{eqnarray}
 \medskip

\noindent {\bf III}
{\it 
 The conformal  transformations  of the correlation functions
(\ref{Lcorrelators})
are generated by the energy--momentum tensor:}
\begin{eqnarray}
\label{LTtrans}
\delta_{\epsilon}T^{\rm\scriptscriptstyle L}(z)
&=&- {1\over 2\pi i}\oint dw\, \epsilon(w) T^{\rm\scriptscriptstyle L}(w)
T^{\rm\scriptscriptstyle L}(z)   \ ,\\
\label{LPtrans}
\delta_{\epsilon,\tilde\epsilon}P(z,\bar z) &=&
-{1\over 2\pi i}\oint \,d w\,
 \epsilon(w) T^{\scriptscriptstyle L}(w)P(z,\bar z) \\
 \nonumber
&& - {1\over 2\pi i}\oint \,d \bar w\, \tilde\epsilon(\bar w)
\widetilde T^{\scriptscriptstyle L}(\bar w)P(z,\bar z) \ .
\end{eqnarray}
Using (\ref{TTOPE}) and (\ref{TPOPE}) they can  be cast in the form
of the conformal Ward identities (CWI):
\begin{eqnarray}
\label{Tward}
\left\langle \,T(w)
\prod_r P(z_r,\bar z_r)\,\right\rangle^{\!\!\!\rm \scriptscriptstyle L}
&=&\sum_r
\left( {Q^2\over 4(w-z_r)^2} +{1\over w-z_r}{\partial\over \partial z_r}\right)
\left\langle \,
\prod_r P(z_r,\bar z_r)\,\right\rangle^{\!\!\!\rm \scriptscriptstyle L} \ ,\\
\label{TTward}
\left\langle \,T^{\rm\scriptscriptstyle L}(u)T(w)
\prod_r P(z_r,\bar z_r)\,\right\rangle^{\!\!\!\rm \scriptscriptstyle L}
&=&
{1+6Q^2\over 2(u-w)^4 }\left\langle \,
\prod_r P(z_r,\bar z_r)\,\right\rangle^{\!\!\!\rm \scriptscriptstyle L} \\
\nonumber
&+& \left({2 \over (u-w)^2} + {1\over u-w}{\partial\over\partial w}\right)
\left\langle \,T(w)
\prod_r P(z_r,\bar z_r)\,\right\rangle^{\!\!\!\rm \scriptscriptstyle L} \\
\nonumber
&+&\sum_r
\left( {Q^2\over 4(u-z_r)^2} +{1\over u-z_r}{\partial\over \partial z_r}\right)
\left\langle \,T(w)
\prod_r P(z_r,\bar z_r)\,\right\rangle^{\!\!\!\rm \scriptscriptstyle L}
\ .
\end{eqnarray}
Apart from the complications related with the non-invariant vacuum and the lack 
of operator--state correspondence
the conditions listed above are exact counterparts of almost all
fundamental  properties of standard CFT.  
So are their consequences. Using CWI one can for instance  reduce
an arbitrary correlation function to the correlation function of punctures alone. 
As in the standard CFT the form
of three puncture correlator is determined up to a constant  
\begin{equation}
\label{3_puncture}
 \left\langle \, P(z_1,\bar z_1)P(z_2,\bar
z_2)P(z_3,\bar z_3)\,\right\rangle = {C\over |z_1
-z_2|^{Q^2\over 2}|z_1 -z_3|^{Q^2\over 2}|z_2 -z_3|^{Q^2\over 2} }\ .
\end{equation}
The familiar state--operator
correspondence can be replaced by the state--puncture
correspondence defined by
\begin{eqnarray}
\label{statepuncture}
L_{-n_1}\dots
L_{-n_N} |\,0\,\rangle^{\rm
\scriptscriptstyle L}
 & \longrightarrow  &
 {\cal L}_{-n_1}\dots
 {\cal L}_{-n_N}
 \cdot
 P(z,\bar z)
 \\
 & \equiv &
 {1\over (2\pi i)^N} \oint_{C_1} dz_1
 {T(z_1)\over (z_1-z)^{n_1-1}}
 \dots
 \oint_{C_N} dz_N{T(z_N)\over (z_N-z)^{n_N-1}}
 P(z,\bar z)\nonumber
 \end{eqnarray}
where the contours  of integration are chosen such that $C_i$ surrounds
$C_{i+1}$ for $i=1,\dots,N-1$, and $C_{N}$ surrounds  the point $z$.
Using this prescription one can associate to each state
${\xi\otimes\bar\xi} \in {\cal V}_h\otimes \bar {\cal V}_h$ ($h={Q^2\over 4}$)
a uniquely determined object
$V_0(\xi,\bar \xi | z,\bar z)$ which we shall call 
the vertex puncture corresponding to $\xi\otimes \bar \xi$.

The only fundamental property of standard CFT we have not yet required is the 
puncture--puncture product expansion (PPPE). 
With our choice of the space of states  there is only one conformal family of 
vertex punctures and the PPPE would be necessarily of the form
\begin{equation}
\label{PPPE_d}
 P(x,\bar x)P(0,0) = Cx^{-h}\bar x^{-\bar h}
\left[P(0,0) + \beta_1 x\ {\cal L}_{-1}\cdot
P(0,0) + \bar\beta_1\bar x\ \bar {\cal L}^{\rm \scriptscriptstyle
L}_{-1}\cdot P(0,0) + \ldots\right]\ ,
\end{equation}
where $C$ is the constant appearing in the three puncture partition
function (\ref{3_puncture}) and the descendant vertex punctures
${\cal L}_{-1}\cdot P(0,0),\ $  $\bar {\cal L}^{\rm
\scriptscriptstyle L}_{-1}\cdot P(0,0),\; $ etc. are defined by
(\ref{statepuncture}).
Using PPPE one could in principle reduce all $N$-puncture correlators
to the 3-puncture functions. The consistency conditions of this procedure yields 
in the case of 4-puncture functions
the bootstrap equation \cite{Belavin:1984vu}. One of its consequences in standard CFT is the
restriction on  possible central charge, conformal dimensions, and
fusion rules known as Vafa's condition \cite{Vafa88}. Using
Lewellen's  derivation of this condition \cite{Lewellen89} one can
show that the PPPE (\ref{PPPE_d}) cannot be satisfied.

According to the properties of the longitudinal sector derived so far 
the theory is completely determined once all $N$-puncture correlators
are known
\begin{equation}
\label{Pcorrelators}
\left\langle \,
\prod\limits_{r=1}^N P(z_r,\bar z_r)\,\right\rangle^{\!\!\!\rm \scriptscriptstyle L} \ .
\end{equation}
It should be stressed
that the correlation functions (\ref{Pcorrelators}) can not be expressed as
the vacuum expectation values of some chronological product of local
operators acting in the Hilbert space 
${\cal V}_h\otimes \bar {{\cal V}}_h$ ($h={Q^2\over 4}$) of the free theory. 
Indeed if it were possible
it would imply the PPPE which is excluded by Vafa's condition. 
This may rise the question on the quantum mechanical meaning 
of the functions (\ref{Pcorrelators}).
In the present paper we adopt the interpretation advocated in 
the geometric approach to the Liouville
theory \cite{Takhtajan:yk} where the $N$-puncture correlators are
understood as partition functions on $N$-punctured spheres
(or in general on higher genus punctured surfaces).
The theory we are looking for can then be seen as a theory of only one local field (the energy--momentum tensor)
in different noncompact 2-dim geometries.

Concluding this section let us briefly comment on the unitarity problem. 
The light-cone approach used in \cite{Hadasz:2002gk} to derive 
the properties  discussed above
incorporates the idea of 
joining--splitting interactions by the assumption that there exists a well defined 
2-dimensional theory on light-cone diagrams. The standard proof of unitarity in this formulation
 is based on the  identification of the light-cone amplitudes as terms of a Dyson
perturbative expansion in the space of multi-string states \cite{kaki,gw}. 
In order to interpret the
integration over moduli of punctured sphere as the integration over interaction times
one uses the Mandelstam map to transform the theory back onto an
appropriate family of light-cone diagrams. 
If the string degrees of freedom are described by standard CFT (as they are in the transverse sector) 
one can use 
the OPE to factorize the light-cone amplitude 
on the free spectrum at any moment between interactions. This in order allows to
interpret the amplitude in terms of interaction vertices and free propagation
yielding the required Dyson expansion structure \cite{Hadasz:2002gk}.
Since in the longitudinal sector the PPPE does not hold  such interpretation is not 
possible and the standard proof of unitarity breaks down.

\section{DOZZ solution}
\setcounter{equation}{0}

The correlators of   operators from the  $({Q^2\over 4},{Q^2\over 4})$ 
conformal family calculated in the DOZZ theory provide a set of functions  satisfying all the requirements listed in the previous section. 
We shall demonstrate that this solution leads to the cut singularities in the non-critical string amplitudes.
The DOZZ structure constants $C(\alpha_3,\alpha_2,\alpha_1),$ 
which determine the three point correlation function 
\[
G^{\rm\scriptscriptstyle L}_{\alpha_1,\alpha_2,\alpha_3}(z_1,z_2,z_3)=
|z_{12}|^{2\gamma_3}|z_{32}|^{2\gamma_1}|z_{31}|^{2\gamma_2}
C(\alpha_3,\alpha_2,\alpha_1)
\]
where
$\gamma_1=\Delta_{\alpha_1}-\Delta_{\alpha_2}-\Delta_{\alpha_3},
\gamma_2=\Delta_{\alpha_2}-\Delta_{\alpha_3}-\Delta_{\alpha_1}, 
\gamma_3=\Delta_{\alpha_3}-\Delta_{\alpha_1}-\Delta_{\alpha_2},$ 
are given by \cite{Teschner:2001rv}
\begin{eqnarray}
\label{3pointsZZ}
&&
\hspace*{-1cm}
C(\alpha_3,\alpha_2,\alpha_1)\; = \; 
\left(\left\lbrack \pi^2 \mu\tilde\mu 
\gamma\left(b^{2}\right)\gamma\left(1/b^2\right)\right\rbrack^{Q/2}
\left(b^{2}\right)^{\frac1b - b}\right)^{Q-\alpha_1-\alpha_2-\alpha_3}  \\
&&
\frac{
\Upsilon_0\Upsilon_b(2\alpha_1)\Upsilon_b(2\alpha_2)\Upsilon_b(2\alpha_3)
}
{
\Upsilon_b(\alpha_1+\alpha_2+\alpha_3-Q)
\Upsilon_b(\alpha_1+\alpha_2-\alpha_3)
\Upsilon_b(\alpha_1+\alpha_3-\alpha_2)
\Upsilon_b(\alpha_2+\alpha_3-\alpha_1)}. \nonumber
\end{eqnarray}
The special function $\Upsilon_b(x)$ has an
integral representation convergent in the strip $0<\Re\,x<Q$
\begin{eqnarray}
&&\log\Upsilon_b(x)\; =\; \int_{0}^{\infty}\frac{dt}{t}
\left\lbrack\left(\frac{Q}{2}-x\right)^{2}e^{-t}-
\frac{\sinh^{2}(\frac{Q}{2}-x)\frac{t}{2}}{\sinh\frac{bt}{2}\sinh\frac{t}{2b}}
\right\rbrack, \nonumber
\end{eqnarray}
$\Upsilon_{0}={\rm res}_{x=0}\frac{d\Upsilon_b(x)}{dx},\;$ $\gamma(x)
= \frac{\Gamma(x)}{\Gamma(1-x)}$ and  the ``dual'' cosmological constant
$\tilde \mu$ is related to $\mu$ by
\begin{equation}
\label{mu_dual}
\left\lbrack \pi\tilde\mu \gamma\left(1/b^2\right)
\right\rbrack^b
= 
\left\lbrack \pi\mu \gamma\left(b^2\right)
\right\rbrack^{1/b}.
\end{equation}
Eq. (\ref{3pointsZZ}) is self-dual: it remains unchanged under $b \to
1/b.$ Its form agrees with the three-point coupling constant proposed in
\cite{Zamolodchikov:1995aa} in the weak coupling region of $b \in
\mathbb{R}.$   If one assumes that the
product $\mu\tilde\mu$ remains real and positive in the strong
coupling region, $b = {\rm e}^{i\theta},$ $\theta \in \mathbb{R},$ (it
is straightforward to check that (\ref{mu_dual}) admits such solution),
then (\ref{3pointsZZ}) defines analytic continuation of
the coupling constant to this region
respecting the self-duality condition, which is sufficient for $C$ to
be real there.

The four-point function for primary fields can be written as the $s$-channel
integral
\begin{eqnarray}
\label{four_point}
G^{\rm\scriptscriptstyle L}_{\alpha_4,\alpha_3,\alpha_2,\alpha_1}(z,\bar{z})
\; = \; 
\int_{\mathcal{D}}\!d\alpha
\; C(\alpha_4,\alpha_3,\alpha)C(\bar\alpha,\alpha_{2},\alpha_{1})
\left|\mathcal{F}_{\alpha}
\left[^{\alpha_3\, \alpha_2}_{\alpha_4\, \alpha_1}\right]
(z)\right|^{2}
\end{eqnarray}
where the  conformal block 
$\mathcal{F}_{\alpha}
\left[^{\alpha_3\, \alpha_2}_{\alpha_4\, \alpha_1}\right]
(z)$ is 
represented  by power series of the form
\begin{eqnarray}
\label{series}
\mathcal{F}_{\alpha}
\left[^{\alpha_3\, \alpha_2}_{\alpha_4\, \alpha_1}\right]
(z)
\; = \; 
z^{\Delta_{\alpha}-\Delta_{\alpha_1}-\Delta_{\alpha_2}}
\sum_{n=0}^{\infty}z^n
\mathcal{F}_{\alpha}
\left[^{\alpha_3\, \alpha_2}_{\alpha_4\, \alpha_1}\right](n).
\end{eqnarray}
For $\alpha_i= \frac{Q}{2}$ 
the set $\mathcal{D}$ coincides with the spectrum 
$\mathbb{S} = \frac{Q}{2} + i\mathbb R^{+}$ which we shall parameterize with
$\frac{Q}{2} + iP.$ The coefficients
$\mathcal{F}_{\alpha}
\left[^{\alpha_3\, \alpha_2}_{\alpha_4\, \alpha_1}\right](n)$  
are rational functions of $\Delta_{\alpha} = \frac{Q^2}{4} + P^2$ with
poles located (for $1 < c < 25$) outside $\mathbb{S}.$
For $b = {\rm e}^{i\theta}$   the formula (\ref{3pointsZZ}) implies
\begin{equation}
\label{C_square}
C\left({Q}/{2},Q/2,Q/2+iP\right)
C\left(Q/2-iP,Q/2,Q/2\right)
 \; = \; 
q\ {\rm e}^{2H_\theta(P)},
\end{equation}
where $q$ does not depend on $P$ and 
\[
H_\theta(P)
\; = \; 
\int\limits_0^\infty\!\frac{dt}{t}
\left[\cos^2\theta\ {\rm e}^{-t} 
+
\frac{1 - 8\sin^2 \frac{Pt}{2} - \cos (2Pt)\cosh\left(t\cos\theta\right)}
{\cosh\left(t\cos\theta\right) - \cos\left(t\sin\theta\right)}\right]
\; \propto \; -P^2 
\]
for $P^2 \gg 0.$ 
The 4-string amplitude for the tachionic external states 
can be written as \cite{Hadasz:2002gk}
\begin{equation}
 \label{tachions_1} 
A \; =\; (2\pi)^{d\over 2} [\alpha]^{-{1\over 2}}
\prod\limits_{\mu = 0}^{d-1} \delta\left(\sum\limits_{r=1}^4 p_r^\mu \right)\;
\int\limits_{\mathbb{C}}\!d^2 x\;
|z|^{\frac{p_1\cdot p_2}{2\alpha}} |1-z|^{\frac{p_2\cdot p_3}{2\alpha}}
G^{\scriptscriptstyle L}(z,\bar z)  \ ,
\end{equation}
where 
$G^{\scriptscriptstyle L}(z,\bar z) =
G^{\rm\scriptscriptstyle
L}_{\frac{Q}{2},\frac{Q}{2},\frac{Q}{2},\frac{Q}{2}}(z,\bar{z}) $ and $p_r$ denote external momenta satisfying the on-mass-shell
condition 
${p^2\over 4\alpha}={d-1\over 12}.$ Using (\ref{C_square}) we write
\begin{eqnarray}
\label{integral}
{\cal I} & \equiv & \int\limits_{\mathbb{C}}\!d^2 z\;
|z|^{\frac{p_1\cdot p_2}{2\alpha}} |1-z|^{\frac{p_2\cdot p_3}{2\alpha}}
   G^{\scriptscriptstyle L}(z,\bar z)  \nonumber \\
&= & 
q\int\limits_{\mathbb{C}}\!d^2 z\;
|z|^{\frac{p_1\cdot p_2}{2\alpha}} |1-z|^{\frac{p_2\cdot p_3}{2\alpha}}
\int\limits_{0}^{\infty}\!dP\; {\rm e}^{2H_\theta(P)}
\left|\mathcal{F}_{\frac{Q}{2} + iP}
\left[^{\frac{Q}{2}\, \frac{Q}{2}}_{\frac{Q}{2}\, \frac{Q}{2}}\right]
(z)\right|^{2}.
\end{eqnarray}
If one can change the order of integration, then --- expanding the
integrand in the power series of $z,\,\bar z$ and integrating 
in the vicinity of $z = 0$ one gets
\begin{equation}
\label{poles}
{\cal I} \; \sim \; \sum_n 
\int\limits_0^{\infty}\!dP 
\frac{d_{n,n}(P)}{\frac{s}{4\alpha} + \frac{Q^2}{2} + 2n + 2P^2 - 2}
{\rm e}^{2H_\theta(P)}
\end{equation}
up to terms analytic in the Mandelstam variable $s=(p_1 + p_2)^2$ at 
the finite part of the complex plane.
$d_{n,n}(P)$ are rational functions of $P$ without poles on the real
axis. 
The presence of poles on
the integration contour for $s < -4\alpha\left(\frac{Q^2}{2}  + 2P^2 -
2\right)$ leads to cuts in ${\cal I}$ in this region.

The cut structure of the scattering
amplitude can be confirmed by choosing a different route in calculating
(\ref{integral}). Taking into account the  convergence 
of the series  (\ref{series}), the  properties of its coefficients 
$\mathcal{F}_{\frac{Q}{2} + iP}
\left[^{\frac{Q}{2}\, \frac{Q}{2}}_{\frac{Q}{2}\, \frac{Q}{2}}\right]\!
(n)$  and of $H_\theta(P)$ one can 
change the order of integration and
summation 
in
calculating the four-point function 
$G^{\scriptscriptstyle
L}(z,\bar z)$ 
in the vicinity of $z=0$. Integrals of the form
\[
\int\limits_0^{\infty}\!dP \; |z|^{2P^2}\ {\rm e}^{2H_{\theta}(P)}\;
\mathcal{F}_{\frac{Q}{2} + iP}
\left[^{\frac{Q}{2}\, \frac{Q}{2}}_{\frac{Q}{2}\, \frac{Q}{2}}\right]\!
(m)\ 
\mathcal{F}_{\frac{Q}{2} - iP}
\left[^{\frac{Q}{2}\, \frac{Q}{2}}_{\frac{Q}{2}\, \frac{Q}{2}}\right]\!
(n)
\]
that arise in this procedure contain terms which for $z \to 0$ behave like
powers of $1/\log|z|.$ Inserted into (\ref{integral}) they 
produce cuts in the complex $s$ plane.
 
\section{Geometric bootstrap hypothesis}
\setcounter{equation}{0}

We shall start with some remarks on the spectrum of conformal weights.
In  the weak coupling regime $c>25$ 
we are interested in 
correlators of local operators with conformal weights satisfying the Seiberg bound $\Delta < {Q^2\over 4}$.
If the theory is coupled to conformal matter such operators are necessary for the  gravitational dressing.
They are supposed to correspond to microscopic, non-normalizable states with imaginary Liouville momenta
\cite{sei}. 
In the geometric approach
they are described by conical (elliptic) singularities with the opening angle $\nu$ related to the imaginary Liouville
momentum $P$ by 
$$
\Delta= \frac{Q^2}{4} + P^2 \; = \;  \frac{Q^2}{4}\left(1-\left(\frac{\nu}{2\pi}\right)^2\right)
$$ 
These operators do not appear  when  the Liouville theory is regarded
as the model for the longitudinal  string excitations in physical dimensions $1<c<25$. 
Indeed the derivation of the light-cone amplitudes
shows that  only the parabolic singularities (limiting case of the conical singularity with
the opening angle $\nu=0$) are relevant for the description of external states. 
The states with imaginary Liouville momenta are also not expected in  factorization
of puncture correlators \cite{gm,sei}.

In order to describe the factorization one has to extend  the geometric approach 
to surfaces with finite holes. 
We assume that the Liouville action and the space of metrics in the path integral
are chosen in such a way that 
the classical solution corresponds to the hyperbolic metric with the
curvature $-\mu<0$ and
the geodesic boundary. We also assume that the  conformal weight of the hole 
is 
$$
\Delta_\ell = \frac{Q^2}{4}\left(1 + \frac{\mu \ell^2}{8\pi^2}\right)
$$
where $\ell$ is the length of the hole circumference  measured with
respect to the metric with constant  
negative curvature $R=-\mu$.  It  depends only on the conformal
class of metrics over which the path integral is taken. 
This assumption is motivated by the properties of the energy--momentum 
tensor of the classical hyperbolic solutions on the cylinder and on sphere
with holes (black-hole solutions in 3-dim gravity). The puncture corresponds to the limiting 
case of the hole with zero circumference. 
Since in the case of puncture
the classical conformal dimension does not receive quantum corrections one may expect 
that this is so for the holes as well (up to the renormalization of the cosmological
constant $\mu$).

Let us now consider the simplest case of 4-puncture correlator.
In the geometric approach it is given by the path integral over the conformal
class of metrics  with parabolic singularities at
puncture locations. Since the conformal weight of the puncture is fixed 
 the space of metric fluctuations   at each puncture is  completely described
by ${\cal L}_n$ operators defined in (\ref{statepuncture}) and coincides
with the space of longitudinal excitations of 
the free string. 
The Liouville action and the space of metrics in the  path integral are
chosen in such a way that for each configuration  
of punctures
one gets a unique  classical solution corresponding to the hyperbolic
metric with scalar curvature $-1.$  
There are three closed geodesics $\Gamma_s,\Gamma_t,\Gamma_u$ in this
geometry separating the punctures into 
pairs (12,34), (13,24), (14,23), respectively.
Let us  cut the
path integral open along the geodesic $\Gamma_s$
 dividing $S$ into the spheres $S_{12},S_{34}$ with one hole and two
punctures ($S=S_{12}\cup S_{34},$ 
$\Gamma_s=S_{12}\cap S_{34}$). 
The classical solution $g$ on $S$ determines classical solutions
$g_{12}$ on $S_{12}$ and $g_{34}$ on $S_{34}.$ 
The initial path integral factorizes into a path integral over the conformal class of $g_{12}$,
 a path integral over the conformal class of $g_{34}$ 
and the integration  over all possible intermediate states.
According to our assumptions the holes on $S_{12}$ and  $S_{34}$
have the same conformal dimension $\Delta_s=
 \frac{Q^2}{4}\left(1 + \frac{\mu \ell_s^2}{8\pi^2}\right)
$ uniquely determined by  the
length $\ell_s$ of the common boundary. It follows that factorization in
each channel involves exactly one conformal family. 
Its  conformal weight depends on the channel and the moduli of the surface.
This is in contrast with 
the DOZZ description  of the weak coupling 
regime where the factorization is independent both of the channel and of the moduli and
involves integration over continuous spectrum of conformal families.

Before we formulate the consistency conditions of the geometric factorization 
introduced above 
some comments on the hole-state correspondence are in order. First of
all since the length of the geodesic 
boundary can assume any positive value one needs a continuous spectrum of conformal weights
$\Delta \geqslant {Q^2\over 4}$ of intermediate states. This however does not necessary mean that
the space of states has to be extended. 
The crucial observation is that 
the geometric factorizaton on exactly one conformal family allows to  interpret the conformal 
weight as a characteristic of the energy--momentum tensor behavior
around the hole 
rather than a characteristic of intermediate states attached to its boundary.
Such interpretation is consistent with our choice of the space of free string states.
Indeed for the central charge  in the range $1\leqslant c\leqslant 25$, 
the left Verma module $ {\cal V}_{Q^2\over 4}$ can be 
realized as the Fock space ${\cal F}^{\rm \scriptscriptstyle L}$ generated
out of the vacuum state $\omega$
by the oscillators 
$$
\begin{array}{rlllll}
{[c_m,c_n]} &=&  m\delta_{m,-n}\ ,
 &\;\;\; m,n\in \mathbb{Z} \setminus\{0\}
\end{array}
$$
 with 
the Virasoro generators given by \cite{Fairlie}
\begin{eqnarray}
L_{n} &=&
 {\textstyle\frac{1}{2}}\sum_{k\neq 0,-n}\!
 :\!c_{-k}c_{n+k}\!: \;+\;i{Q\over \sqrt{2}} n c_n \;+\;
 {Q^2\over 4}\delta_{n,0}
 \nonumber \ .
\end{eqnarray}
For any real positive number $P$ (the Liouville momentum)
one can construct on ${\cal F}^{\rm \scriptscriptstyle L}$ 
the local operator 
$$
T_{\scriptscriptstyle P}(z)=
\sum\limits_{n}L_n^{\scriptscriptstyle P} z^{-n-2} \;\;\;,
\;\;\;L_n^{\scriptscriptstyle P}\;=\;\left\{
\begin{array}{lll}
L_n^{\scriptscriptstyle L} + 2Pc_n &{\rm for}& n\neq 0\\
L_0^{\scriptscriptstyle L} + {P^2} &{\rm for}& n= 0
\end{array}
\right.
$$ 
satisfying 
\begin{eqnarray}
\nonumber
T_{\scriptscriptstyle P}
(w)T_{\scriptscriptstyle P}(z)
&=&{\frac12 (1+6Q^2)\over (w-z)^4 }
+ {2 \over (w-z)^2}T_{\scriptscriptstyle P}(z)
+{ 1\over w-z}\partial T_{\scriptscriptstyle P}(z)
 + \dots \ , \\
\label{Tsingularity}
T_{\scriptscriptstyle P}(z)
\Omega
&=& {\Delta\over z^2}
\Omega
+ {1\over z}L^{\scriptscriptstyle P}_{-1}
\Omega
+ {\rm regular \; terms}\ . 
\end{eqnarray}
where $\Delta = {Q^2\over 4}+{P^2}$.
With an appropriate choice of the coordinates around the hole, the state-hole correspondence 
takes the form
\begin{eqnarray}
\label{statehole}
L^{\rm \scriptscriptstyle P}_{-n_1}\dots
L^{\rm \scriptscriptstyle P}_{-n_N} |\,0\,\rangle^{\rm
\scriptscriptstyle L}
 & \longrightarrow  &
 {\cal L}_{-n_1}\dots
 {\cal L}_{-n_N}
 \cdot
 H_\ell(z,\bar z)
 \\
 & \equiv &
 {1\over (2\pi i)^N} \oint_{C_1} dz_1
 {T(z_1)\over (z_1-z)^{n_1-1}}
 \dots
 \oint_{C_N} dz_N{T(z_N)\over (z_N-z)^{n_N-1}}
 H_\ell(z,\bar z)\ .\nonumber
 \end{eqnarray}
As in the case of punctures  one can associate to each state
${\xi\otimes\bar\xi\in{\cal F}^{\rm \scriptscriptstyle L}\otimes \bar{{\cal F}}^{\rm \scriptscriptstyle L}}$
a uniquely determined object
$V_\ell^{\rm \scriptscriptstyle L}(\xi,\bar \xi | z,\bar z)$ which we shall call 
the vertex corresponding to to the state $\xi\otimes \bar \xi$ applied at the 
hole of the circumference $\ell$.
We have assumed  that the location of the hole can be defined 
as a point at which the analytic continuation of the energy--momentum tensor is singular.

Although the correlators of punctures, holes, and in general, vertexes
are not realized as vacuum expectation values 
of local operators all their conformal properties are 
exactly the same as in the standard CFT. 
In particular  an arbitrary three vertex correlator is determined up to
the structure constant 
$C(\ell_1,\ell_2,\ell_3)$. It takes the form
\begin{eqnarray*}
& &\hspace{-60pt}\left\langle \, 
 V^{\rm \scriptscriptstyle L}_{\ell_1}(\xi_1,\bar \xi_1 |z_1,\bar z_1)
V^{\rm \scriptscriptstyle L}_{\ell_2}(\xi_2,\bar \xi_2 |z_2,\bar z_2)
V^{\rm \scriptscriptstyle L}_{\ell_3}(\xi_3,\bar \xi_3 |z_3,\bar z_3)\,\right\rangle \;= \;C(\ell_1,\ell_2,\ell_3) \\
& \times &{\rho^{\ell_1\,\ell_2\,\ell_3}(\xi_1,\xi_2,\xi_3)\over 
(z_1 -z_2)^{\Delta_1+\Delta_2-\Delta_3}
(z_1 -z_3)^{\Delta_1+\Delta_3-\Delta_2}
(z_2 -z_3)^{\Delta_2+\Delta_3-\Delta_1} }\\
&\times &
{\rho^{\ell_1\,\ell_2\,\ell_3}(\bar\xi_1,\bar\xi_2,\bar\xi_3)\over 
(\bar z_1 -\bar z_2)^{\bar\Delta_1+\bar\Delta_2-\bar\Delta_3}
(\bar z_1 -\bar z_3)^{\bar\Delta_1+\bar\Delta_3-\bar\Delta_2}
(\bar z_2 -\bar z_3)^{\bar\Delta_2+\bar\Delta_3-\bar\Delta_1} }
\end{eqnarray*}
where $\Delta_i = \Delta_{\ell_i} +|\xi_i|$, $\bar\Delta_i = \Delta_{\ell_i} +|\bar\xi_i|$.
The  trilinear forms $\rho^{\ell_1\,\ell_2\,\ell_3}(\xi_1,\xi_2,\xi_3)$, universal for all CFT,
are uniquely determined by CWI \cite{Teschner:2001rv}.
Also the notion of the conformal block 
remains unchanged
\begin{eqnarray*}
{\cal F}_{Q,\,\ell}\left[
^{\ell_3\,\ell_2}_{\ell_4\,\ell_1 }  
\right]\!(z)&=& z^{\Delta_\ell -\Delta_{\ell_2} -\Delta_{\ell_1}}
\sum\limits_{n=0}^{\infty} z^n {\cal F}_{Q,\,\ell}\left[
^{\ell_3\,\ell_2}_{\ell_4\,\ell_1 }  
\right]\!(n)\\
{\cal F}_{Q,\,\ell}\left[
^{\ell_3\,\ell_2}_{\ell_4\,\ell_1 }  
\right]\!(n) &=& \sum\limits_{I,J \in {\cal I}_n} 
\rho^{\ell_4\,\ell_3\,\ell}(\omega,\omega,\xi_I)B_{Q,\ell}^{IJ}(n)
\rho^{\ell\;\;\ell_2\,\ell_1}(\xi_J,\omega,\omega) \ .
\end{eqnarray*}
(In the formulae above $\omega$ denotes the vacuum state in ${\cal
F}^{\scriptscriptstyle L},$
$\left\{ \xi_I \right\}_{I\in {\cal I}_n}$ is a basis on the level $n$ in ${\cal F}^{\scriptscriptstyle L}$,
and $B_{Q,\ell}^{IJ}(n)$ is the inverse to the Gram matrix on the level $n$ in the Verma module with
the central charge $ c=1 +6Q^2$ and the highest weight 
$\Delta_\ell = \frac{Q^2}{4}\left(1 + \frac{\mu \ell^2}{8\pi^2}\right)$.)

In the simplest case of the four puncture correlator, 
\[
G^{\scriptscriptstyle L}(x,\bar x)
\; = \; \lim\limits_{z,\bar z\to \infty} z^{Q^2\over 2}\bar z{}^{Q^2\over 2}
 \left\langle
  P(z,\bar z)P(1,1)P(x,\bar x)P(0,0)
  \right\rangle\ ,
\]
the geometric factorization in the $s$-channel reads
\begin{equation}
\label{factorization}
G^{\scriptscriptstyle L}(x,\bar x)= C(0,0,\ell_s(x,\bar x))C(\ell_s(x,\bar x),0,0) 
\left|{\cal F}_{Q,\,\ell_s(x,\bar x)}\!\left[^{\,0\;0\,}_{\,0\;0\,}
\right]\!(x)\right|^2\ ,
\end{equation}
where $\ell_s(x,\bar x)$ is the length of the closed geodesic $\Gamma_s$. Let us note that 
 the amplitude does not factorizes into a holomorphic
and an anti-holomorphic part. 

The formula (\ref{factorization}) involves the length of closed
geodesic in the $s$-channel as a function of the fourth  puncture
location $x$. In the vicinity of $x=0$  it can be expanded in powers of 
$1/\log|x|$ \cite{Wolf}. If we assume that 
the structure constants are real analytic at $\ell_i=0$ then the geometric factorization
(\ref{factorization}) yields the same behavior of the Liouville correlator
and the same analytic structure of the string amplitude 
as in the case of the DOZZ solution.

The basic consistency condition of the presented approach is 
the obvious requirement that the geometric factorization in each channel yields the
same result. In the slightly more general case of the four hole correlator it leads to the 
geometric bootstrap equations:
\begin{eqnarray*}
&&\hspace{-50pt} C(\ell_4,\ell_3,\ell_s)C(\ell_s,\ell_2,\ell_1)
\left|{\cal F}_{Q,\,\ell_s}\!\left[^{\ell_3\,\ell_2}_{\ell_4\,\ell_1 }  \right]\!(x)\right|^2\\
&=&
C(\ell_4,\ell_1,\ell_t)C(\ell_t,\ell_2,\ell_3)
\left|{\cal F}_{Q,\,\ell_t}\!\left[^{\ell_1\,\ell_2}_{\ell_4\,\ell_3 }  \right]\!(1-x)\right|^2\\
&=&
|x|^{-4\Delta_{\ell_2}} C(\ell_1,\ell_3,\ell_u)C(\ell_u,\ell_2,\ell_4)
\left|{\cal F}_{Q,\,\ell_u}\!\left[^{\ell_3\,\ell_2}_{\ell_1\,\ell_4 }  \right]
\!\left({\textstyle {1\over x}}\right)\right|^2
\end{eqnarray*}
where $\ell_s, \ell_t, \ell_s$ are the lengths of the closed geodesics in corresponding channels.
As in the standard CFT one can promote the geometric bootstrap equations to 
the basic dynamical principle of the theory.

\section{Discussion and conclusions}
\setcounter{equation}{0}

Whether the geometric bootstrap equations 
provides a plausible solution of the Liouville theory 
depends on our ability to calculate the
structure constants. Due to the complicated non-linear nature of the equations 
a direct analysis of this problem
is prohibitively difficult. It seems that the only strategy available is to construct a
candidate for the structure constants and then to verify whether it satisfies
the geometric bootstrap equations.

The simplest possibility is to consider the DOZZ structure constant. 
This proposal might be motivated by the existence of the analytic continuation of the 3-point 
function from the weak to the strong coupling region and from the elliptic to
the hyperbolic weights. Another problem is to verify a slightly stronger hypothesis that
not only the structure constants but also the correlators
of the geometric approach and of  the DOZZ theory are identical. In the case of 4-point correlators it takes
the form of the following equation
\begin{eqnarray*}
C(\ell_4,\ell_3,\ell_s(x,\bar x))C(\ell_s(x,\bar x),\ell_2,\ell_1) 
\left|{\cal F}_{Q,\,\ell_s(x,\bar x)}\!\left[^{\ell_3\, \ell_2}_{\ell_4\, \ell_1}\right]\!(x)\right|^2 
\;\;\;\;\;\;\;\;\;\;\;&&\\
 \;\;\; \propto\; 
\int\limits_{0}^\infty\!d\ell
\; C(\ell_4,\ell_3,\ell)C(\ell,\ell_2,\ell_1) 
\left|\mathcal{F}_{Q,\,\ell}
\left[^{\ell_3\, \ell_2}_{\ell_4\, \ell_1}\right]
(x)\right|^{2}&& \;.
\end{eqnarray*}

The second possible way to find the structure constants is to analyze a path integral
representation of the 3-point function. It requires an appropriate
generalization of the Liouville 
action of the geometric approach from the conical and the puncture singularities to finite
holes. It involves in particular the open problem
of auxiliary parameters
and Polyakov's conjecture in the case of hyperbolic singularities. 
This topic seems to be especially interesting as it would provide a way for 
the semi-classical analysis of the theory by methods analogous to that 
developed by L.\ Takhtajan in the case of the puncture singularities.

Even if the structure constants were known, the verification of the
geometric bootstrap equations would still be a challenging task.
First of all the equations involve the lengths of closed geodesics in each
channel as functions of the locations of punctures. Up to our knowledge
even in the simplest case 
of 4-punctured sphere such functions are not known \cite{Hempel,Wolf}. Secondly the conformal block  
function is also not known in this case and can be studied only by the
numerical methods developed in  
 \cite{Zamolodchikov:ie,Zam}. Both problems are of their own interest
going beyond the present context.

Let us finally comment on the problem of consistent interactions of non-critical Nambu-Goto 
string. 
As one could expect from the continuous spectrum the DOZZ approach
leads to the tree string amplitudes with cut singularities. 
The same result was obtained in a more speculative way in the geometric approach.
It follows that the Liouville theory description of the longitudinal
dynamics applied in the framework of the standard string perturbation
expansion does not lead to a consistent  interacting non-critical Nambu-Goto string. 
This leaves almost no room to maneuver for constructing such a theory. 
 
\section*{Acknowledgements}

We would like to thank J\"{o}rg Teschner and B{\'e}n{\'e}dicte Ponsot 
for discussion. Z.J. is grateful to Michael Wolf
for the correspondence and bringing to our attention the reference
\cite{Wolf}. The work of L.H. was supported by the EC IHP
network HPRN-CT-1999-000161. Laboratoire de Physique Th{\'e}orique is
Unit{\'e} Mixte du CNRS UMR 8627.

\end{document}